\documentclass{article}
\usepackage{cite}
\usepackage{amsmath,amssymb,amsfonts}
\usepackage{algorithmic}
\usepackage{graphicx}
\usepackage{textcomp}
\usepackage{xcolor}
\usepackage[utf8x]{inputenc}
\usepackage{tikz}
\usepackage{pgfplots}
\usepackage{xspace}
\usepackage{listings}
\usepackage[list=true]{subcaption}
\usepackage{url}
\usepackage{booktabs}
\usepackage{multirow}
\pgfplotsset{compat=newest}
\usetikzlibrary{matrix,graphs,plotmarks,quotes,positioning,calc}

\usepackage{color}
\usepackage{xcolor}
\usepackage{paralist}

\definecolor{pblue}{rgb}{0.13,0.13,1}
\definecolor{pgreen}{rgb}{0,0.5,0}
\definecolor{pred}{rgb}{0.9,0,0}
\definecolor{pgrey}{rgb}{0.46,0.45,0.48}
\newcommand\tardis{\textit{Tardis}\xspace}

\date{}
\def\BibTeX{{\rm B\kern-.05em{\sc i\kern-.025em b}\kern-.08em
		T\kern-.1667em\lower.7ex\hbox{E}\kern-.125emX}}
\begin{document}
	
	\title{Combining Dynamic Symbolic Execution, Machine Learning and Search-Based Testing to Automatically Generate Test Cases for Classes}

	\author{Matteo Modonato\\Dipartimento di Informatica,
				Sistemistica e Comunicazione\\
			Università di Milano Bicocca\\
			Milano, Italy \\
			m.modonato1@campus.unimib.it}

	\maketitle
	
	\begin{abstract}
	\noindent This article discusses a new technique to automatically generate test cases for object oriented programs.
At the state of the art, the problem of generating adequate sets of complete test cases has not been satisfactorily solved yet.
There are various techniques to automatically generate test cases (random testing, search-based testing, etc.) but each one has its own weaknesses.
This article proposes an approach that distinctively combines dynamic symbolic execution, search-based testing and machine learning, to efficiently generate thorough class-level test suites.
The preliminary data obtained carrying out some experiments confirm that we are going in the right direction.
	\end{abstract}
	

\section{Introduction}


%
Automatically generating  class-level test cases for object oriented programs, e.g., programs in Java, requires to identify sequences of method calls that comply with the legal APIs of the programs under test, and suite to thoroughly exercise the classes in the programs.
\emph{Complete test cases} shall further include all method calls needed to properly instantiate and set any relevant non-primitive input that participates in the test cases. The inputs can be objects of either the same type as the class under test or types specified in other classes, including library classes, or yet structures of multiple interconnected objects of many classes.

At the state of the art, the problem of generating adequate sets of complete test cases has not been satisfactorily solved yet. 
On one hand, many existing approaches systematically characterise thorough sets of test-relevant input objects, 
based on either functional or structural testing criteria,
but provide no support for generating method sequences to instantiate complete test cases with these inputs~\cite{Edvardsson:SurveyTestDataGeneration:CCSE:1999,Korel:automated:TSE:1990,McMinn:SBSESurvey:STVR:2004,Braione:SymbExeHeapInputs:ESECFSE:2015,Tillmann:pex:TAP:2008,Boyapati:Korat:ISSTA:2002}.
On the other hand, the approaches that succeed in producing complete test cases may often miss many relevant test objectives that depend on rare input states or complex input objects~\cite{Xie:Symstra:TACAS:2005,Pacheco:Randoop:ICSE:2007,Thummalapenta:seeker:OOPSLA:2011,Fraser:Evosuite:TSE:2013,Braione:SUSHI:ISSTA:2017,Tillmann:pex:TAP:2008}.


Existing test generators build complete test cases based on random testing, search-based testing and symbolic execution~\cite{Pacheco:Randoop:ICSE:2007,Fraser:Evosuite:TSE:2013,Xie:Symstra:TACAS:2005,Tillmann:pex:TAP:2008,Thummalapenta:seeker:OOPSLA:2011,Braione:SUSHI:ISSTA:2017}. Random testing samples the possible method sequences and their inputs uniformly, but may miss test objectives that depend on singular or quasi-singular inputs~\cite{Pacheco:Randoop:ICSE:2007}. 
Search-based testing explores the space of the possible test cases with meta-heuristic search strategies, 
guided by fitness functions that encode the relevant test objectives, e.g., the branches on the code~\cite{Fraser:Evosuite:TSE:2013,Xie:Symstra:TACAS:2005}. It can outperform random testing in several cases, 
but may still miss many relevant test objectives that induce flat fitness functions and local optima that hamper the convergence of the search strategies.

Other approaches embrace symbolic execution to precisely characterize the execution conditions of the program paths that exercise the test objectives, and then identify complete test cases by satisfying such execution conditions~\cite{Tillmann:pex:TAP:2008,Thummalapenta:seeker:OOPSLA:2011,Braione:SUSHI:ISSTA:2017}. 
While symbolic execution may foster precise deductions on how to reach the test objectives, it suffers more than random and search-based testing in coping with  programs that maintain implicit invariants of the validity of the input objects, e.g., by controlling the visibility and the accessibility of class variables. The ignorance of the invariants makes symbolic execution waste most testing budget in trying to instantiate infeasible execution paths that depend on (unforeseen) invalid input objects.

This paper introduces \tardis, a unit test generator for Java programs that 
distinctively combines dynamic symbolic execution, search-based testing and machine learning, to efficiently generate thorough class-level test suites. 
The approach of \tardis consists of 
\begin{inparaenum}[(i)]
\item exploring the path space of the target programs with dynamic symbolic execution, 
\item instantiating complete test cases with a genetic search algorithm guided with fitness functions that represent the satisfiability of the symbolic path conditions, and 
\item prioritizes the symbolic formulas that more likely correspond to feasible program paths based on an original classification algorithm trained on the characteristics of the formulas for which it ether succeeds or fails overtime.   
\end{inparaenum}


This paper is organized as follows. Section~\ref{sec:motivating} describes the existing approaches based on a working example, and highlights their limitations that motivate our work on \tardis.
Section~\ref{sec:tardis:testgen} presents the \tardis algorithm. Section~\ref{sec:tardis:classifier} explains in further detail the core technical novelty of \tardis, that is, the path section strategy  based on machine learning. Section~\ref{sec:evaluation} discusses initial results on the effectiveness of \tardis. Section~\ref{sec:conclusions} 
provides our conclusions and outlines our research plans.


\section{Motivating Example and Related Work}
\label{sec:motivating}
In this section we discuss the current test generation generation approaches that, to the best of our knowledge, can generate class-level complete test cases for object oriented programs. We highlight in particular the approaches that target Java classes.

Listing~\ref{CodeExample} introduces a sample program that we use in this section to discuss the strengths and the limitations of the test generation approaches at the state of the art, and lately in the next section 
to describe the \tardis algorithm.
This program includes a branch that is difficult to test both with purely dynamic test generators, like random and search-based test generators~\cite{Pacheco:Randoop:ICSE:2007,Fraser:Evosuite:TSE:2013}, and
with the existing test generators based on symbolic execution and dynamic symbolic execution~\cite{Tillmann:pex:TAP:2008,Thummalapenta:seeker:OOPSLA:2011,Anand:JPFSE:TACAS:2007,Braione:SUSHI:ISSTA:2017}. The challenging branch is the one that leads to execute line~\ref{sample:challenging} of method \texttt{SampleClass.run}, where the program returns the string \texttt{"Yes"}. Executing line~\ref{sample:challenging} requires a test case that does not traverse 
either line~\ref{sample:assign:k} or line~\ref{sample:assign:flag} while executing through the loops in the program: At line~\ref{sample:assign:k} the program assigns \texttt{k=1}, and thus executing this line causes the violation of the assertion at line~\ref{sample:assertion} and in turn the termination of the program before line~\ref{sample:challenging}; At line~\ref{sample:assign:flag} the program assigns \texttt{flag=false}, and thus executing this line leads to skip the execution of line~\ref{sample:challenging}.  

\begin{lstlisting}[caption={Code example},label={CodeExample},language=Java, escapechar=&] 
public class SampleClass {
	private final int[] a;
	private final int[] b;
	
	public SampleProgram (
	        int a0, int a1, int a2, int a3, int a4, 
			int b0, int b1, int b2, int b3, int b4, 
			int b5, int b6, int b7, int b8, int b9) {
		a = new int[] { a0,  a1,  a2,  a3,  a4};
		b = new int[] { b0,  b1,  b2,  b3,  b4,  b5,  b6,  b7,  b8,  b9};
		for (int i = 0; i < b.length; i++) {
		    if (b[i] < 0) throw new RuntimeException(); &\label{sample:invariant}&
		}
	}
	
	public String run() {
		int k = 0;
		boolean flag = true;
		for (int i = 0; i < A.length; i++) {&\label{sample:loop1}&
			if (A[i] > i * 1000) {&\label{sample:if1}&
				for (int j = 0; j < b.length; j++) {&\label{sample:loop2}&
					if (b[j] < -A[i]) { /* This branch is infeasible */&\label{sample:if2}& 
					    k = 1; &\label{sample:assign:k}&
					}
				}
			} else {
				flag = false; &\label{sample:assign:flag}&
			}
		}
		assert (k == 0);&\label{sample:assertion}&
		if (flag) {
			return "Yes"; &\label{sample:challenging}&
		} else {
			return null;
		}
	}
}
\end{lstlisting}

We refer to the sample program of Listing~\ref{CodeExample} to explain the strengths and exemplify the limitations of the main approaches that address the generation of test cases for Java classes at the state of the art, aiming to motivate our work on \tardis to overcome these limitations. In particular, we discuss approaches based on random testing (e.g.,~\cite{Pacheco:Randoop:ICSE:2007}), search-based testing (e.g.,~\cite{Fraser:Evosuite:TSE:2013}), and both classic and dynamic symbolic execution (e.g.,~\cite{Anand:JPFSE:TACAS:2007, Braione:SUSHI:ISSTA:2017,Tillmann:pex:TAP:2008,Thummalapenta:seeker:OOPSLA:2011}).

\paragraph{Random Testing}
Random testing generates test cases by sampling the legal method sequences and the corresponding input values at random.
The tool Randoop is popular test generator based on random testing that addresses Java classes~\cite{Pacheco:Randoop:ICSE:2007}.

The main advantage of using random testing is that it can readily generate many complete test cases based on the legal APIs of the program under test.
For instance, for the class \texttt{SampleClass } in Listing~\ref{CodeExample}, which specifies only a class constructor and the method \texttt{run}, it is straightforward for a tool like Randoop to find legal method sequences that consist of calling the class constructor with random values as inputs, and then executing method \texttt{run}. For example, Randoop may easily generate a the test case like

\begin{lstlisting}[language=Java, numbers=none] 
SampleClass s = new SampleClass(123, 456, ...);
s.run();
\end{lstlisting}

On the other hand, random testing can hardly identify test cases that execute the parts of the program under test that depend on almost singular inputs or combinations of inputs. For example, Randoop would hardly generate a test case that executes the branch at line~\ref{sample:challenging} of our sample program, because, to this end, it should set the first five inputs of the class constructor to a combination of values that make the if-statement at line~\ref{sample:if1} execute only the then-branch, and the subsequent ten inputs of the class constructor to non-negative values. In the sample program, the first five inputs of the constructor determine the values in the array \texttt{a} that the if-statement at line~\ref{sample:if1} sequentially evaluates at each pass. The other ten inputs of the constructor must be non-negative values to avoid the exception at line~\ref{sample:invariant}.
Thus,  assuming for simplicity that Randoop would randomly pick with equal probability values that do or do not satisfy the condition of the if-statement, and positive or negative values, respectively, Randoop has \(1/2^{15}\) probability to pick ten satisfying values in a row, that is, about 0.003\% chances out of a value space of \(64^{15}\) if the integer data range on 64 bits.

\paragraph{Search-based Testing}
Search-based testing guides the selection of random test cases with meta-heuristic algorithms based on fitness functions that capture the execution conditions of test objectives. Thus, search-based testing can mitigate the issues of random testing that derive in the first place from picking inputs uniformly without being unaware of the relation between the inputs and the test objectives, e.g., executing the branches our sample program. The tool Evosuite is a popular search-based test generator for Java classes~\cite{Fraser:Evosuite:TSE:2013}.

Evosuite can be parametrized with several code-based fitness functions, but we limit our discussion to the specific fitness functions by which EvoSuite  addresses the branches in the code. In this specific setting, 
EvoSuite will consider a fitness function
for each then- and else-branch of 
 each conditional statement in the program under test, e.g., the then-branch and else-branch of  if-statement at line~\ref{sample:if1} of our sample program.

The algorithm of EvoSuite proceeds as follow. It starts with a randomly generated population of test cases, and scores these test cases according all the fitness functions. 
For instance, let us assume that the initial population includes a test case like

\begin{lstlisting}[language=Java, numbers=none] 
SampleClass s = new SampleClass(-5, -5, -5, -5, -5, 0, ...\* all zeros *\);
s.run();
\end{lstlisting}

\noindent that executes the else-branch of the if-statement at line~\ref{sample:if1}, but does never execute the then-branch of this if-statement. Thus, Evosuite  scores this test case as optimal with respect to the fitness function of the else-branch, and sub-optimal for the then-branch.
Then, while progressing in the search, EvoSuite records the test cases that satisfy some target branch for the first time, and restricts its attention to the fitness functions of the yet-unsatisfied branches to produce test suites that execute as many branches as possible.
In our example, Evosuite marks the fitness function of the else-branch as satisfied, records the test case, and proceeds in the optimizing the test cases with respect to the  fitness function of the then-branch.
With respect to this latter fitness function, EvoSuite scores the test case based on the minimal change that is required to make the condition of the branch evaluate to true. For instance with the above test case, Evosuite observes that the condition of the branch evaluates as $-5>0$ (which is false), and thus a change of $6$ units of the left operand in the comparison would suffice for the condition to evaluate to true.  While continuing through its (genetic) algorithm, Evosuite will then favour this test case to survive in the evolving population, until finding other test cases with better fitness scores, e.g, a test case that misses the considered true-branch of 5, 4 or less units, until eventually producing a satisfying test case with respect to this condition.

Unfortunately, despite the specific focus on  program branches may often lead EvoSuite to outperform random testing in code coverage for many programs, EvoSuite can hardly cope with the hard branch at line~\ref{sample:challenging} of our sample program. In fact, the condition of this branch depends on a boolean variable and thus all test cases that miss this branch have the same fitness score of $1$ (since a change of a single boolean unit would suffice to hit the branch) resulting in a local-optimum of the fitness landscape. 
Thus, EvoSuite has no guidance for executing  the branch at line~\ref{sample:challenging}, and thus 
it falls back to having exactly the same low probability of random testing of finding a satisfying solution for this branch.

\paragraph{Symbolic Execution}
Symbolic execution identifies the execution conditions of the program paths, which it refers to as the \emph{path conditions}, and generates test cases by solving the path conditions to concrete values. 
For instance, by analyzing the program paths that lead to the hard branch at line~\ref{sample:challenging} of our sample program, we can compute the path condition
\(A \ne null \wedge A.length > 0 \wedge A[0] > 0 \wedge \dots \wedge A[1] > 1000 \wedge \dots \wedge A[2] > 2000 \wedge \dots \wedge A[3] > 3000 \wedge \dots \wedge A[4] > 4000 \wedge \dots\), where the symbol $A$ denotes the input array associated to the class variable \texttt{SampleClass.a} and the dots mark further conditions that we omitted for the sake of simplifying the presentation, e.g., the conditions on the items in the array \texttt{SampleClass.b}. To solve this path condition, we have to assign the five items in the array \texttt{SampleClass.a} with increasingly larger positive numbers, which is the requirement to execute the branch at line~\ref{sample:challenging}.
The tools JPF~\cite{Anand:JPFSE:TACAS:2007} and JBSE~\cite{braione:enhancing:FSE:2013,Braione:SymbExeHeapInputs:ESECFSE:2015} are test generators for Java classes based on symbolic execution.

Symbolic execution can outperform the purely dynamic approaches of random testing and search-based testing for identifying test cases that execute specific program paths, since, as in the above example, it models the execution conditions of the paths with precise path-condition formulas. However, symbolic execution
 suffers more than the purely dynamic technique in i) computing test cases based on legal method sequences, and ii) efficiently coping with programs with many possible execution paths. 

We first discuss the difficulty of symbolic in computing test cases based on legal method sequences. As we discussed above, by solving a path condition of the sample program, symbolic execution can compute an assignment of the array \texttt{SampleClass.a} that allows for executing the hard branch in the program. However, there is not straightforward way for symbolic execution to understand which legal sequence of method calls can set these values into an instance of class \texttt{SampleClass}. Indeed, understanding that the values in the array depend on the first five parameters of the constructor is not obvious, and the required method sequences and program APIs can easily be much more complicated in other programs than in this example.

Traditional symbolic executors, including JPF and JBSE, circumvent this issue by breaking the restrictions designed in the program APIs, e.g., 
by augmenting the programs under test with setter methods for all class attributes, or using the Java Reflection APIs to access the memory at low level. However, this is arguably an undesirable solution for many reasons. On one hand it results in test cases that are not in developer-format and can be difficult for the developers to understand and maintain. On the other hand, it may result in illegal test cases and false alarms. For example, with reference to the sample program, the symbolic executor JBSE can characterize a program path that violates the assertion at line~\ref{sample:assertion} with a path condition that includes the sub-formula \(A[0] > 0 \wedge B[0] < -A[0]\), then compute a solution in which the first item in array \texttt{SampleClass.b} holds a negative value, and finally produce a test case that enforces a negative item in \texttt{SampleClass.b[0]} by relying on the Java Reflection APIs, thus demonstrating the violation of the assertion. Unfortunately this test case is in fact a false alarm, since the legal constructor of the class, which JBSE is in fact bypassing, does not allow for setting negative values in the array \texttt{SampleClass.b}.

The other difficulty of symbolic execution is to cope with programs with many or infinitely many paths, since analyzing large sets of paths can be  inefficient, especially if it is difficult to spot the infeasible paths. 
For example, in our sample program, there are only 32 feasible paths that correspond to the combinations of the true and false alternatives of the if-statement at line~\ref{sample:if1}, while the program scans through the array \texttt{SampleClass.a}. However, since symbolic execution is unaware of the invariant, the class constructor maintain on the items  in the array \texttt{SampleClass.b}, which can only hold positive values due to the exception raised at line~\ref{sample:invariant},
the alternatives of the  if-statement at line~\ref{sample:if1} further combine
with the $2^{10}$ combination of alternatives of the if-statement at line~\ref{sample:if2} on the items in the two arrays. Thus, for our sample program, a test generator based on symbolic execution like JBSE would end up with raising many false alarms, while missing most feasible paths, including the path to execute the hard branch in the program. 

SUSHI is an approach that addresses the limitations of symbolic execution by computing test cases based on legal method sequences~\cite{Braione:SUSHI:ISSTA:2017}. SUSHI converts the path conditions computed with JBSE into fitness functions with which SUSHI can steer EvoSuite to compute test cases that satisfy the path conditions. Thus SUSHI does not produces false alarms, since EvoSuite does not yield test cases for the path conditions that cannot be satisfied  with legal test cases. However, SUSHI pays high performance penalties with programs with many infeasible paths, as in the case of our example, since it wastes a lot of time in waiting for EvoSuite to expire the inconclusive search attempts for the unsatisfiable path conditions. SUSHI can be fed with specifications that encode the properties of the valid inputs, which help SUSHI to identify the unsatisfiable path conditions in advance. Unfortunately, requiring the developers to produce these specifications questions the actual level of automation of the approach.

Dynamic symbolic execution combines symbolic and concrete execution in a feedback loop, with the aim of steering symbolic execution to analyze the path conditions only for the program branches that directly originate from a concretely executed path. In a nutshell, dynamic symbolic works by iterating the following feedback loop: It executes a test case, analyzes the path condition of the corresponding program path, synthesizes the path conditions of the alternative branches by negating all clauses one-by-one (that is, each new path condition negates a single clause of the original one), solves those path conditions to obtain new test inputs, and iterates the process based on the generated test inputs. PEX is a popular test generator that implements dynamic symbolic execution for programs in C++~\cite{Tillmann:pex:TAP:2008}. A benefit of dynamic symbolic execution execution is to abandon as early as possible the analysis of the paths that originate at branches with unsatisfiable path conditions. This can mitigate, though not generally solve, the impact of the infeasible paths on the performance of the analysis. For example, with reference to our sample program, dynamic symbolic execution would still try the many infeasible choices related to the if-statement at line~\ref{sample:if2}, but would not unfold the combinations of sub-paths related to those infeasible choices. 

Several authors propose path selection heuristics to steer dynamic symbolic execution towards uncovered and relevant branches in the code by discriminating the sub-paths that can statically reach the target branches ~\cite{burnim:crest:ASE:2008,Person:DISE:PLDI:2011,Baluda:BidirectionalSymbExe:TSE:2016}. However the results are inconclusive in most cases. For our sample program, a path selection strategy based on targeting the hard branch at line~\ref{sample:challenging} and the uncovered (but infeasible) branch at line~\ref{sample:assign:k} would not bring any particular benefit.

\section{Test Generation with \tardis}
\label{sec:tardis:testgen}
\tardis combines symbolic execution, machine learning and search-based testing, to automatically generate test cases that thoroughly exercise Java classes with legal method sequences:
\begin{itemize}
\item \tardis exploits symbolic execution in the embodiment known as dynamic symbolic execution~\cite{Godefroid:DART:PLDI:2005} to precisely identify the execution conditions of the program paths (i.e., the path conditions), with respect to both the primitive and the structured inputs of the programs under test~\cite{khurshid:generalized:tacas:2003,Braione:JBSE:ESECFSE:2016,Anand:JPFSE:TACAS:2007}.
\item \tardis embraces machine learning to define a path selection heuristic based on classifying
the path conditions that are either likely satisfiable or likely unsatisfibale, according to the characteristics of both the satisfiable and unsatisfiable path conditions that \tardis learns during the test generation process. 
\item \tardis satisfies the symbolic conditions selected as above in the fashion of search-based testing, that is, by incrementally optimising a population of legal method sequences with respect to fitness functions that suitably represents the satisfaction of the path conditions~\cite{Fraser:Evosuite:TSE:2013,Braione:SUSHI:ISSTA:2017}. 
\end{itemize}


\begin{figure}[h]
    \centering
    \includegraphics[width=12cm]{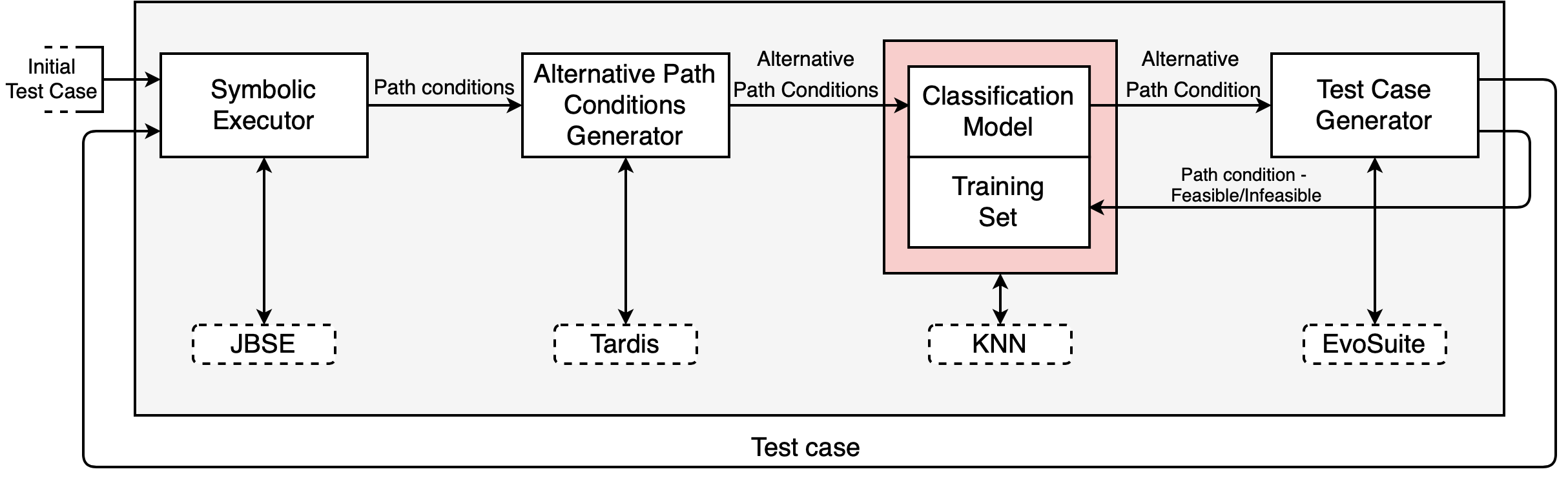}
    \caption{Tardis workflow}
    \label{fig:TardisWorkflow}
\end{figure}

Figure~\ref{fig:TardisWorkflow} illustrates the workflow of \tardis that consists of iterating through three main steps: i) dynamic symbolic execution, ii) path selection and iii) test case generation.

\paragraph{Dynamic Symbolic Execution Step}
The \emph{dynamic symbolic execution} step consists in executing an available test case, computing the path condition of the program path that the test case executes, and synthesizing alternative path conditions by negating the conjunctive clauses of the path condition obtained with the test case. The considered test case can be either a test case that is initially available from external sources (e.g., given to \tardis as input), or a test case that \tardis itself generated at a previous iteration. 

For instance, with reference to the sample program of Listing~\ref{CodeExample}, let us assume that \tardis executes the dynamic symbolic execution step by based on a test case like

\begin{lstlisting}[language=Java, numbers=none] 
SampleClass s = new SampleClass(0, 0, ... /* all zeros */);
s.run();
\end{lstlisting}

\noindent \tardis proceeds as follow. First, by executing this test case, \tardis identifies the program path that iterates 5 times the through the loop at line~\ref{sample:loop1}, executes the else-branch at line~\ref{sample:assign:flag} at all iterations, and terminates by raising the violation of the assertion at line~\ref{sample:assertion}. Then, by analyzing this execution path, \tardis computes the path condition \(A \ne null \wedge A.length > 0 \wedge A[0]\le 0 \wedge A.length > 1 \wedge A[1]\le 1000 \wedge A.length > 2 \wedge A[2]\le 2000 \wedge A.length > 3 \wedge A[3]\le 3000 \wedge A.length > 4 \wedge A[4]\le 4000 \wedge A.length \le 5\); where the symbol $A$ denotes the input array associated to the class variable \texttt{SampleClass.a}. Next, by negating each of the eleven clauses of this path condition, \tardis can synthesize the following alternative path conditions in which we highlighted the negated clauses:
\begin{itemize}
\item \(\bf A.length \le 0\)
\item \(A.length > 0 \wedge \bf A[0]> 0\)
\item \(A.length > 0 \wedge A[0]\le 0 \wedge \bf A.length \le 1\)
\item \(A.length > 0 \wedge A[0]\le 0 \wedge A.length > 1 \wedge \bf A[1]> 1000\)
\item \(A.length > 0 \wedge A[0]\le 0 \wedge A.length > 1 \wedge A[1]\le 1000 \wedge \bf A.length \le 2\)
\item \(A.length > 0 \wedge A[0]\le 0 \wedge A.length > 1 \wedge A[1]\le 1000 \wedge A.length > 2 \wedge \bf A[2]> 2000\)
\item \(A.length > 0 \wedge A[0]\le 0 \wedge A.length > 1 \wedge A[1]\le 1000 \wedge A.length > 2 \wedge A[2]\le 2000 \wedge \bf A.length \le 3\)
\item \(A.length > 0 \wedge A[0]\le 0 \wedge A.length > 1 \wedge A[1]\le 1000 \wedge A.length > 2 \wedge A[2]\le 2000 \wedge A.length > 3 \wedge \bf A[3]> 3000\)
\item \(A.length > 0 \wedge A[0]\le 0 \wedge A.length > 1 \wedge A[1]\le 1000 \wedge A.length > 2 \wedge A[2]\le 2000 \wedge A.length > 3 \wedge A[3]\le 3000 \wedge \bf A.length \le 4\)
\item \(A.length > 0 \wedge A[0]\le 0 \wedge A.length > 1 \wedge A[1]\le 1000 \wedge A.length > 2 \wedge A[2]\le 2000 \wedge A.length > 3 \wedge A[3]\le 3000 \wedge A.length > 4 \wedge \bf A[4]> 4000\)
\item \(A.length > 0 \wedge A[0]\le 0 \wedge A.length > 1 \wedge A[1]\le 1000 \wedge A.length > 2 \wedge A[2]\le 2000 \wedge A.length > 3 \wedge A[3]\le 3000 \wedge A.length > 4 \wedge A[4]\le 4000 \wedge \bf A.length > 5\)
\end{itemize}
\tardis feeds these path conditions to its next step, the path selection strategy, that aims to choose on
which path conditions 
\tardis will allocate test generation attempts.

\paragraph{Path Selection Step}
In  \emph{path selection step} \tardis aims to discriminate between the likely satisfiable and likely unsatisfiable path conditions, and prioritize the test generation attempts with respect to former ones. In fact, during the test generation process, \tardis will iterate the dynamic symbolic execution step for all the test cases that it incrementally generates, and will thus keep on synthesizing increasingly more additional path conditions, possibly including many unsatisfiable path conditions that correspond to infeasible paths in the program under test. For instance, continuing with the above example, the six path conditions that \tardis synthesized by negating conditions that predicate on the length of the array \texttt{a} are unsatisfiable, since the constructor of class \texttt{SampleClass} maintains the invariant that  \texttt{a} always contains exactly five items. Namely, these path conditions are the first, third, fifth, seventh, ninth and eleventh in the list outlined above.
As we already commented, allocating too many test generation attempts on unsatisfiable path conditions may strongly penalize the overall efficiency of the test generator, wasting most computational time in attempts that cannot converge. 


The core of the path selection step of \tardis
is a classification model based on machine-learning that assigns selection priorities to the path conditions, aiming to assign low priority to the path condition that it can predict as likely infeasible.  \tardis  builds and incrementally refines the classification model based on
the evidences in support of either satisfiable or unsatisfiable formulas that it collects 
from the path conditions met
during the test generation process. In fact, the path conditions that \tardis computes by analyzing available and generated test cases contribute supporting evidences of satisfiable formulas, while the path conditions on which \tardis allocated unsuccessful test generation attempts contribute supporting evidences of formulas that might be unsatisfiable. For instance, by analyzing the initial test case exemplified above, \tardis collects evidence that the formulas that comprise the path condition computed for the test case are indeed satisfiable, e.g., the formulas \(A.length > 0\), \(A[0]\le 0\), \(A.length > 1\) and so forth. While, if lately \tardis would fail in the attempt to generate a test case for the path condition synthesized by negating the third clause, i.e., the path condition \(A.length > 0 \wedge A[0]\le 0 \wedge \bf A.length \le 1\),
it can thereon hypothesize that the formula that corresponds to the new clause in this path condition \(A.length \le 1\) might be unsatisfiable. 

\tardis exploits the incrementally collected supporting evidences for engineering a classifier based on the classic KNN (K-nearest neighbours) machine learning algorithm. The classifier computes the similarity  of the formulas in the not-yet-selected path conditions with respect to the satisfiable and unsatisfiable formulas that Tradis collected as supporting evidences, and classifies the path conditions according to the type of supporting evidences that are  most similar to them. For instance, after 
collecting the supporting evidence that the formula \(A.length < 1\) might be unsatisfiable as described above, \tardis  would classify as likely unsatisfiable the other path conditions in which the last clause is \(A.length < \langle CONSTANT\rangle\), because these clauses look similar to \(A.length < 1\). Namely, these are the first, fifth, seventh and ninth path condition in the list outlined above. Based on this classification results, \tardis might thus assign low selection priorities to these unsatisfiable path conditions. Section~\ref{sec:tardis:classifier} discusses the details on how \tardis collects the supporting evidences by identifying the sub-formulas that represent independent satisfiability problems in the path conditions (a step called path condition slicing in other papers~\cite{Cadar:KLEE:OSDI:2008,Aquino:Recal:ISSTA:2015}), and computes the similarity and priority scores.

The priorities that Tradis assigns to the path conditions based on the results of the KNN-based classifier induce a probabilistic path selection strategy, in which the path conditions with high priority scores are selected with higher probability than the path conditions with low priority scores. In this way, \tardis favours the allocation of testing attempts on the likely satisfiable path conditions, but acknowledges that the supporting evidences on unsatisfiable formulas come with some uncertainty. Indeed, the meta-heuristic search algorithm that \tardis uses to generate the test cases may not succeed either because the selected path conditions are truly unsatisfiable or sometimes because it is hard for the algorithm to solve them, despite they are satisfiable. The probabilistic selection leaves open the chance that now and then \tardis may allocate test generation attempts to low-priority path conditions. If any of these attempts eventually succeeds, \tardis will update the supporting evidence according to the corresponding satisfiability results, and reconsider its classifications accordingly.

\paragraph{Test Case Generation Step}
The \emph{test case generation} step \tardis relies on the genetic algorithm implemented in the tool Evosuite~\cite{Fraser:Evosuite:TSE:2013} steered with a fitness function that scores as optimal the test cases that satisfy the target path conditions, SUSHI~\cite{Braione:SUSHI:ISSTA:2017}. The fitness function has the property of yielding increasing better fitness for the test cases that satisfy increasing amounts of clauses of those path conditions. In a nutshell, the test generation algorithm starts with a population of test cases built with legal method sequences selected at random, iterates with producing offsprings of new the test cases obtained by randomly mutating and crossing the current test cases in the style of genetic algorithms, and favors the test cases with greater fitness to survive in the population with higher probability than the ones with lower fitness. Eventually, the algorithm either converges to a solution, i.e., a test case with optimal fitness, or gives up after a given time budget.
For instance, it \tardis selected the path condition \(A.length > 0 \wedge \bf A[0]> 0\), the second in the list outlined above, for a test generation attempt, it synthesize an evaluator program that computes the fitness function based on this path condition, and call Evosuite with this evaluator program as input. As a result, Evosuite may provide a test case like 

\begin{lstlisting}[language=Java, numbers=none] 
SampleClass s = new SampleClass(123, 0, ... /* all zeros */);
s.run();
\end{lstlisting}

\noindent that consists of a legal method sequence satisfies the path condition. 

\tardis will provide this test case as output for the users, and will feed this test case back in its workflow to synthesize and select further alternative path conditions out of it, and generate further test cases that cover other execution paths of the program.  

\section{Path Selection}
\label{sec:tardis:classifier}
This section presents in detail the path selection strategy of \tardis, which is the core technical novelty of the approach.

The set of alternative path conditions generated by \tardis can be large, and thus a path selection problem materializes. Furthermore many alternative path conditions generated by \tardis could be unsatisfiable, that is path conditions for which there is no input that makes executable the associated path and from which no test case can be generated. The more the path selection strategy chooses path conditions, and therefore paths, from which it is not possible to generate new test cases, the more resources are wasted and the process of generating a test suite will be long. Therefore it is essential to be able to "avoid" these unsatisfiable path conditions, i.e. to analyze them with less probability in comparison with the satisfiable ones.

In order to make the path selection technique as efficient as possible and to direct the tool to the choice of feasible paths, we decided to use a machine learning model to classify a-priori the satisfiable path conditions from the unsatisfiable ones, that is, predict whether it is possible or not possible to generate a test case from a particular path condition.
In this way, \tardis aims to solve the problem of waste of resources during the analysis of the unsatisfiable path conditions.

Figure \ref{fig:TardisWorkflow} shows how \tardis   integrates with the classification model, designed to solve the problem. The selection of the alternative path conditions generated by \tardis to be sent to EvoSuite is made by considering the classification (feasible or infeasible) computed by the model (highlighted in red in the figure) for each path condition. After the model classifies the alternative path conditions generated, \tardis more likely chooses those that according to the classifier can lead to the generation of a test case, in order to pass them to EvoSuite.

\tardis receives feedback directly from the test generation attempts that it submits to Evosuite at runtime; therefore the classification model is designed to incrementally adapt its predictions as the training set grows over time. This is because, whenever EvoSuite is able or not to generate a test case from a path condition, this information is added to the training set to be used to classify future path conditions. In other words, whenever EvoSuite is able to generate a test case from a path condition, it is added to the training set with the "feasible" label. On the other hand, if EvoSuite cannot generate a test case, the path condition is added to the training set associated with the "infeasible" label.

\begin{figure}[h]
    \centering
    \includegraphics[width=12cm]{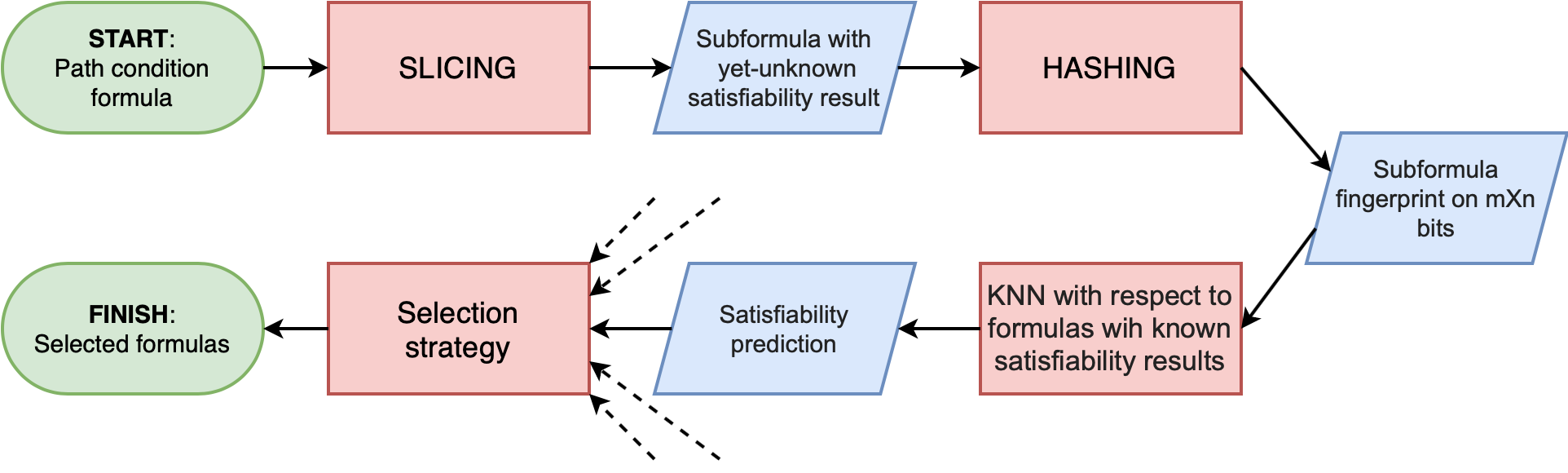}
    \caption{Classification procedure diagram.}
    \label{fig:ProcedureDiagram}
\end{figure}

The designed classification process can be summarized as follows. We use a classification model (KNN) capable of classifying the path conditions as either feasible or infeasible. This occurs by evaluating the similarity of the path conditions still to be analyzed with the path conditions already analyzed for which \tardis knows if they are satisfiable or unsatisfiable.

To make the classification problem less “noisy”, \tardis preprocess the path conditions (Slicing) in order to identify the sub-formulas that contribute, independently of each other, to the unsatisfiability of the path conditions.

A hashing-based methodology to represent path conditions has been defined to foster the efficient calculation of the similarity between the path conditions.

\subsection{Path condition Slicing}

A preprocessing phase of the data has been defined to help the classifier to better identify the patterns characterizing the feasibility and the infeasibility. Since the alternative path conditions generated by \tardis originate from the path conditions of concrete test cases, it is known that the path condition from which the alternative one derives is satisfiable (this is because \tardis uses only path conditions that have been verified as satisfiable as candidates for the generation of alternative path conditions, i.e. those from which it managed to generate a test case). Thus, in order to study the unsatisfiability of a new (alternative) path condition, we need to focus only the last part of the new path condition, that is, the part that changed with respect to the satisfiable path condition, along
with all other parts that predicate on the same variables as the last part.

To do this, a formula manipulation technique called Slicing was used. Formula Slicing can be defined as follows: It slices a formula into sub-formulas that are mutually independent, that is, sub-formulas containing different variables. Slicing splits a formula into a set of independent sub-formulas. Two formulas are independent if and only if they do not share any variable. The satisfiability of a complex formula can be determined from the satisfiability of its sub-formulas: if all the independent sub-formulas are satisfiable, the original formula is satisfiable too, otherwise the original formula is unsatisfiable. Slicing formulas produces smaller formulas that are easier to handle and more likely to be equivalent to previously solved formulas.

The alternative path conditions generated by \tardis can be divided into two parts: a prefix and a suffix. The suffix is the last part of the formula, i.e. the clause that \tardis modifies to generate a new alternative path condition to analyze. The prefix is all that part of the formula that remains unchanged. This part, since it was previously already analyzed by \tardis, is satisfiable by construction (for the reason mentioned above).
%
For this reason, Slicing is performed only relative to the suffix. All conditions of the path condition that are independent from the suffix are satisfiable by construction.

\begin{figure}[h]
    \centering
    \includegraphics[width=8cm]{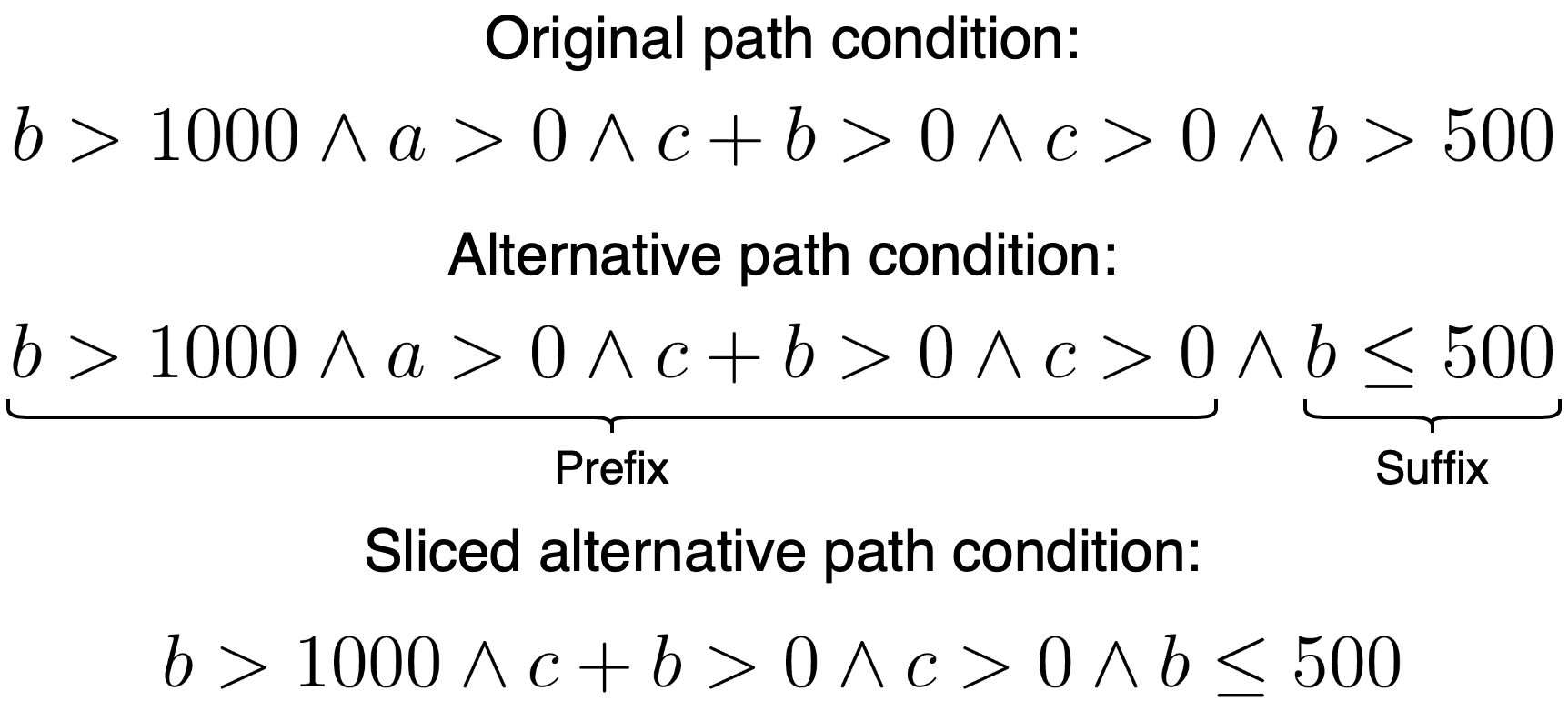}
    \caption{Example of Slicing applied to the \tardis system.}
    \label{fig:Slicing}
\end{figure}

\tardis implements the slicing of the path conditions by considering two types of dependencies: variable-based dependency and object-based dependency.

\paragraph{Variable-based dependency}

Two formulas have a mutual dependency if they predicate on intersecting sets of variables.
In the example shown in figure \ref{fig:Slicing} the first path condition represents a path condition previously analyzed by \tardis from which the tool generated a test case. The second one is a possible alternative path condition generated by the tool: the last condition has been modified. This will be considered as the suffix; all the previous part will be the prefix of the path condition. It is possible to see that in the final formula, after applying Slicing, the condition $a > 0$ is no longer present because this clause does not share any variable with the suffix. The $c > 0$ clause is not directly connected with the suffix but is still kept in the formula. This is because it is indirectly connected by the $c + b > 0$ clause. The transitivity relationship is therefore fundamental when the Slicing procedure is applied.

\paragraph{Object-based dependency}

We also take into consideration the cases where the relationship between different variables that belong to the same object might depend on implicit invariants maintained by the methods that control the instantiation and the access policy of the objects. Consequently, the dependence based on the object to which the variables belong was also taken into consideration, in addition to the dependency between clauses based on the direct or indirect sharing of variables. In the same way as variable-based dependency, object-based dependency is applied transitively.

Let's take as an example a simple path condition like $\{O2.c<0 \land O1.b>0 \land O1.a>0\}$. By applying the same procedure as in figure \ref{fig:Slicing}, the formula is divided into suffix (last clause) and prefix (remaining clauses). Using transitive variable-based dependency only, the three clauses would be independent; however using the concept of objects-based dependency too, we can see how the suffix is related to the clause $O1.b>0$ because they belong to the same $O1$ object. The resulting path condition after the slicing process is $\{O1.b>0 \land O1.a>0\}$.\\

The Slicing procedure in \tardis is a transitive closure of the dependencies between the clauses of a formula in relation to the last clause of the formula itself, where the dependencies are managed according to variables and objects in a transitive way. The idea is based on the fact that, to establish the satisfiability of a formula whose prefix is certainly satisfiable and to which a new part (suffix) is added, the satisfiability depends only on the clauses of the prefix that share, directly or indirectly, variables or objects with the suffix.

\subsection{Formulas fingerprints}\label{section:FormulasFingerprints}

Since path conditions are conjunction of clauses, the objective of the classification model is to recognize the similarity between the path conditions that share sets of joint clauses. In this way, if the model recognizes that a path condition is similar as a subset of conjuncts to one that was previously associated with the infeasible label (the greater the number of clauses they share and the greater their similarity), then the tool will penalize this path condition, in terms of priorities for the transition to EvoSuite. Vice versa, if the model recognizes that a path condition is similar as a subset of conjuncts to one that has previously been associated with the feasible label, it will give higher priority to this one.

To do this, a methodology was defined for the representation of the path conditions generated by \tardis, so that these could be effectively managed by a classification model. It was used a Bloom Filter structure that allows to represent the containment of elements between sets in a compact way; in our specific case, this property was adapted to represent path conditions as a containment of sets of clauses.\\

A Bloom filter is a space-efficient probabilistic data structure that is used to test whether an element is a member of a set. False positive matches are possible, but false negatives are not; in other words, a query returns either "possibly in set" or "definitely not in set." Elements can be added to the set, but not removed; the more items added, the larger the probability of false positives. An empty Bloom filter is a bit array of m bits, all set to 0. There must also be k different hash functions defined, each of which maps an element, by hashing the element itself, on one of the m positions of the array. The number k of hashing functions used is fixed and much smaller than the number of bits m in the array.

To add an element to the structure, we feed it to each of the k hash functions to get k array positions. Then set the bits at all these positions to 1.

To query for an element (test whether it is in the set), feed it to each of the k hash functions to get k array positions. If any of the bits at these positions is 0, the element is definitely not in the set; if it were, then all the bits would have been set to 1 when it was inserted. If all are 1, then either the element is in the set, or the bits have by chance been set to 1 during the insertion of other elements, resulting in a false positive. In this data structure there is no way to distinguish between the two cases.

\begin{figure}[h]
    \centering
    \includegraphics[width=10cm]{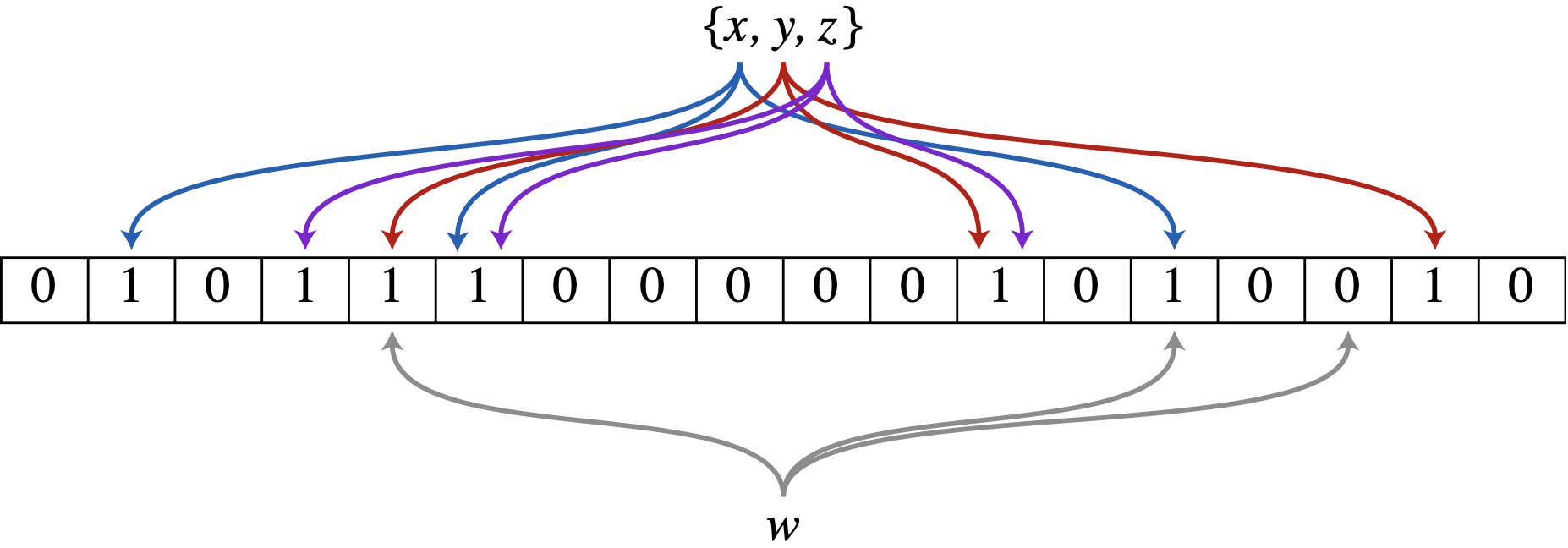}
    \caption{The figure shows an example of the use of a Bloom Filter. The three elements \textit{\{x, y, z\}} are represented in the data structure. The colored arrows show the bits on which the three elements of the set are mapped (in this case three hash functions for each element). The element \textit{w} is not contained within the set \textit{\{x, y, z\}} because one of the three hash functions maps the element on a bit set to 0.}
    \label{fig:BloomFilter}
\end{figure}

Since this data structure is able to represent and manage sets of elements, it was chosen to represent our path conditions produced by \tardis as sets of conditions. Specifically, each path condition is represented through a Bloom Filter structure: each condition of the path condition is inserted into the structure by mapping it to the m bits through k hash functions.

\begin{figure}[h]
    \centering
    \includegraphics[width=12cm]{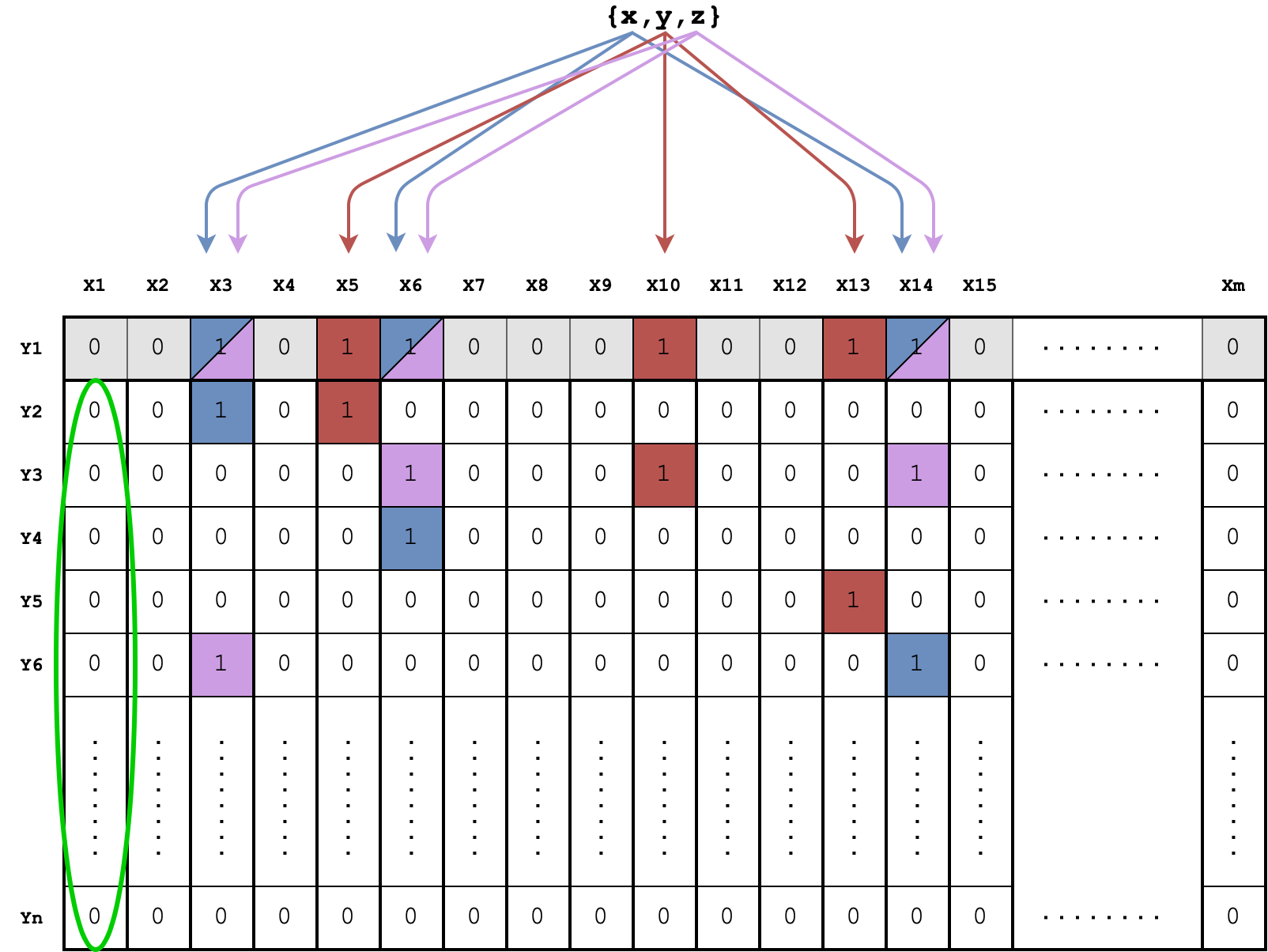}
    \caption{Multidimensional Bloom Filter structure example.}
    \label{fig:MultidimensionalBloomFilter}
\end{figure}

A data structure based on the Bloom Filter, but slightly modified (Multidimensional Bloom Filter: figure \ref{fig:MultidimensionalBloomFilter}), was developed so that it could better adapt to the operation of the \tardis tool.
Using a classic single-row Bloom Filter (figure \ref{fig:BloomFilter}), the problem arises that similar formulas can be mapped to different points in the bit space. This could cause problems to a classification model like our type since it may not always be the same formula, but similar formulas, to cause unsatisfiability. The Multidimensional Bloom Filter structure was created to solve this problem.

This structure works on two dimensions, the X axis and the Y axis. We get a matrix of size m x n. To understand how it works, the matrix can be ideally divided into two main areas: the first row Y1 (gray in figure \ref{fig:MultidimensionalBloomFilter}), and the columns below X1, X2, ..., Xm starting from row Y2. Each bit of the first row corresponds to a column of bits below. For example, the column relating to the bit of the first row X1Y1 is the column of bits starting from position X1Y2 and ending at position X1Yn (circled in green in figure \ref{fig:MultidimensionalBloomFilter}). This organization is applied to all the bits in the array.

This structure was designed to generate a bit organization that could best characterize the similarity between formulas. This factor is important to solve our classification problem since it may not always be the same formula, but similar formulas, to cause unsatisfiability. Specifically, this organization was designed to manage two different types of path conditions: abstract path conditions and concrete path conditions. From a single path condition generated by \tardis, two path conditions are developed. The first is the abstract one in which, within the clauses that compose it, there are no numeric elements such as specific values or indices of the arrays. The second is the concrete one, that is the original one.

Considering a path condition of this type: $\{a[0]>6 \; \& \; b[2]>3 \; \& \; a[5]>6\}$; the concrete clauses will be $\{a[0]>6\}$, $\{b[2]>3\}$ end $\{a[5]>6\}$; the abstract clauses will be $\{a[.]>.\}$, $\{b[.]>.\}$ end $\{a[.]>.\}$

The Multidimensional Bloom Filter structure is used as follows: the clauses of the abstract path conditions are mapped, through hash functions, to the first row of the matrix. The clauses of the concrete path conditions are subsequently mapped on the underlying columns corresponding to the bits set to 1 in the first row. In this way, similar clauses are always mapped in bits positioned in the same area (same columns) and the problem of sparsity, mentioned above, is avoided.

In figure \ref{fig:MultidimensionalBloomFilter} \textbf{x, y, z} represent the abstract clauses of an abstract path condition ($\{a[.]>.\}$, $\{b[.]>.\}$ and $\{a[.]>.\}$ of the example above). After that the abstract clauses are mapped on the first row through the 3 hash functions, the corresponding concrete clauses ($\{a[0]>6\}$, $\{b[2]>3\}$ end $\{a[5]>6\}$ of the example above) of the concrete path condition are mapped on the columns corresponding to the bits set to 1 of the first row. In this example it is possible to notice how the clauses \textbf{x} and \textbf{z} are similar clauses since they turn on the same 3 bits of the first row; this means that they have the abstract clauses that coincide. Consequently, the concrete clauses are mapped on the same 3 columns (pink and blue colored bits). The \textbf{y} clause, on the other hand, is not similar to the other two because its abstract clause is mapped on different bits of the first row. Consequently, the concrete clauses are mapped on the corresponding columns which, however, are different from those of the other two clauses (bits colored red).

In the example in the figure, an optimal case is managed, that is a situation in which each clause of the path condition is mapped on three different bits through the three different hash functions. This fact, due to properties intrinsic to the type of structure, is not always respected; it may happen that different hash functions map different clauses on the same bit. Therefore the possibility of "collisions" between bits is accepted because the elimination of the possibility of duplication of the outputs is not guaranteed (it is possible that the same output value is returned from different hash functions and different input data).

\subsection{Satisfiability predictions}

K-Nearest Neighbors was chosen as the classification model to be integrated into \tardis.

The k-nearest neighbors (KNN) is an algorithm used in patterns recognition to classify objects, it is based on the characteristics of the objects close to the one considered. The predictions are made for each new instance (x) by looking for the most similar K instances (the neighbors) in the entire training set and choosing the output variable through these K instances only. Distance metrics are used to determine which of the K instances in the training set is more like a new input.

When KNN is used to solve a classification problem, the output is calculated as the class with the highest frequency from the most similar K instances (the K neighbors). Basically, each neighbor votes for their own class and the class with the highest number of votes is taken as a prediction.

The main reason we choose this model is that the Knn has training time equal to 0 and this guarantees that it does not have to stop the \tardis execution to retrain the model while the training set gradually grows. This feature is fundamental since the training set, through which the classifications are made, gradually grows while \tardis works.

The distance measurement called Jaccard is used to determine which instances in the training set are the most similar to a new input. This distance is calculated between two $x$ and $y$ vectors consisting of $TRUE, FALSE$ or $0,1$ values. In our case the distance is calculated between two matrices, of size $m X n$, generated by the procedure defined in section \ref{section:FormulasFingerprints}; each matrix represents an alternative path condition generated by \tardis. The Jaccard distance is a distance used to measure the similarity between two $x$ and $y$ vectors. The similarity is returned in output as a value between 0, which indicates a non-similarity between the two vectors, and 1, which means that the two vectors are identical.

For logical vectors $x$ and $y$, we define the following:
\begin{itemize}
\setlength\itemsep{0em}
\item $a11 =$ number of times where $x_i=1$ e $y_i=1$
\item $a10 =$ number of times where $x_i=1$ e $y_i=0$
\item $a01 =$ number of times where $x_i=0$ e $y_i=1$
\item $a00 =$ number of times where $x_i=0$ e $y_i=0$
\end{itemize}
where $i$ indicates the position in the vector. Similarity is calculated using the following formula:
\begin{equation} \label{JaccardDistance}
a11 \over{a11 + a10 + a01}
\end{equation}

For each of the alternative path conditions generated by \tardis the output of this classification model will be label 0 or 1, where:
\begin{itemize}
\setlength\itemsep{0em}
\item \textbf{0} if the BloomFilter structure refers to an unsatisfiable path condition (the classifier predicts that the tool will not be able to generate a test case for that path condition).
\item \textbf{1} if the BloomFilter structure refers to a satisfiable path condition (the classifier predicts that the tool will be able to generate a test case for that path condition).
\end{itemize}

The K value (number of neighbors to be considered for classification) is set to 3.

\subsection{Path selection strategy}

The choice of path conditions to be passed to EvoSuite, considering the classification given by the model, is managed in a probabilistic way. The alternative path conditions which have still to be analyzed are divided according to two elements: classification and voting. Classification is the label, 0 or 1, assigned through the classification model to each of the path conditions still to be analyzed. Voting is a value that measures by which majority the K neighbors voted for the classification (voting can take the value 2 or 3 working with a K equal to 3).

Path conditions are divided into four sets based on the classification obtained via KNN: \textit{label1Voting3, label1Voting2, label0Voting2 and label0Voting3}. 
As mentioned previously, the classification model is an incremental model since the training set grows during tool execution. Whenever EvoSuite is able or not able to generate a test case from a path condition, this information is added to the training set to be used to classify future path conditions. For this reason, the classifications of the path conditions that still have to be analyzed are recalculated several times during the execution of \tardis and may change over time; consequently, the distribution of path conditions within the four sets changes over time too.

The system will pick up the path condition from these sets with following probabilities:
\begin{enumerate}
\setlength\itemsep{0em}
\item label=1, voting=3 : 50\%
\item label=1, voting=2 : 30\%
\item label=0, voting=2 : 15\%
\item label=0, voting=3 : 5\%
\end{enumerate}

Path conditions that are classified as feasible through an unanimous vote by the neighboring k have greater probability of extraction than all the others.

Once the set from which to take the path condition is selected, the cumulative is performed to select the actual path condition to be passed to EvoSuite. Specifically, the cumulative is calculated on the values obtained from the averages of the Jaccard distances of the k neighbors.

The cumulative can be defined in the following way: given a list of values, each element will become the sum between itself and its previous elements. For example, if I have a list composed of the values \{5, 7, 2, 4\} the corresponding cumulative will be \{5, 12, 14, 18\}. The goal is to generate a serie of intervals with the cumulative of the averages of Jaccard distances. For the example above, the ranges are: [0, 5), [5, 12), [12, 14), [14, 18). By randomly extracting a number between the minimum and the maximum value of the intervals, this will belong to one of these. Based on the interval to which the random value belongs, the corresponding path condition will be extracted to be passed to EvoSuite.

Using the cumulative calculated on the averages of the distances of the selected set, the elements with greater distance, and therefore with greater similarity, will be extracted with greater probability because the corresponding intervals will be wider.

Summarizing, two random extractions are carried out to choice the path condition to be passed to EvoSuite; the first is used to select one of the four sets from which to extract the path condition (with the probabilities defined previously); the second random extraction is used to extract the actual path condition and is carried out using the cumulative average of the Jaccard k distances.

\subsection{Incremental learning algorithm}

As already mentioned, the classification model integrated in \tardis is incremental and consequently the training set grows over time. For this reason, the classifications of the path conditions that still have to be analyzed are recalculated several times during tool execution. To ensure that this does not affect performance, a threshold-based cashing mechanism was implemented: the reclassification of the already classified path conditions, which must be carried out again since the training set changed, is carried out every time the training set grows by a certain threshold. Path conditions that have never been classified are classified immediately. In this way continuous reclassification of all path conditions is avoided.

As far as the mechanism for choosing the set from which to take the path condition concerns, if the set selected through the random extraction is empty, the immediately following non-empty set is chosen. For example, if label1Voting3 is extracted but it is empty, then label1Voting2 is chosen if there is at least one element inside the latter; If label1Voting2 is also empty, then we pass to label0Voting2 and so on.

The bloom Filter structure was implemented as a 64 x 16 bit size matrix that uses 3 hash functions to map each clause. This configuration was used for the experiments in the chapter \ref{sec:evaluation}.

\section{Empirical Evaluation}
\label{sec:evaluation}
This section reports on a set of initial experiments that  evaluate the effectiveness of \tardis and in particular of the path selection strategy explained in the previous section. 

The evaluation metric used is based on  comparing the results of \tardis with and without our classification model, to verify whether the former is more efficient in choosing the feasible paths. Therefore we compare the results of \tardis in terms of number of feasible paths explored, when the selection strategy is used, with the case when it is not used. In the second case \tardis uses a queue based on the FIFO (First In, First Out) principle as a method for selecting the paths to be analyzed, i.e. it uses FIFO to manage the transition to EvoSuite of the alternative path conditions generated.

\paragraph{Experimental setting}
The comparison is based on the sample program discussed in section \ref{sec:motivating}.

A \textit{timeBudgetDuration} equal to 30 has been set to carry out this evaluation; this means that the data obtained and analyzed refer to the first 30 minutes of \tardis execution. In addition, the \textit{numOfThreads} parameter has been set equal to 5; this means that 5 threads can run simultaneously and then analyze 5 path conditions simultaneously. The maximum time that can be dedicated to analyze a path condition, beyond which it is established that this is satisfiable, is a configuration parameter and can vary; during these experiments the parameter was set at 180 seconds.

\begin{figure}[h]
    \centering
    \includegraphics[width=12cm]{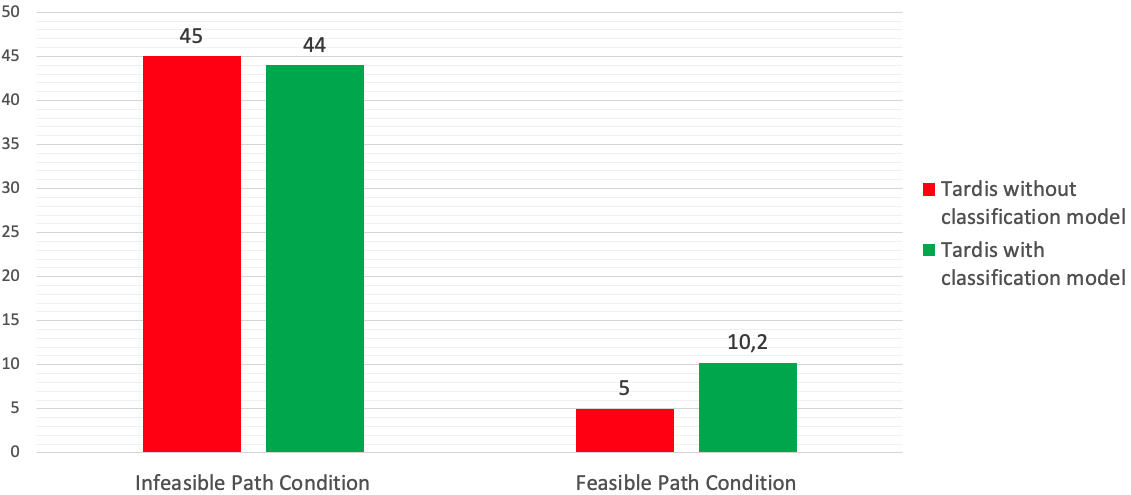}
    \caption{Distribution histogram of the analyzed path conditions.}
    \label{fig:DataHistogram}
\end{figure}

\paragraph{Results}
Image \ref{fig:DataHistogram} shows the number of path conditions analyzed when \tardis uses the classifier (in green) or work without classifier (in red). The displayed data are the average of the data obtained on several \tardis executions. 

Interestingly, the data highlight that the feasible path conditions analyzed during the \tardis executions with the classifier doubled in comparison to those analyzed during the execution of the tool without classifier. At the same time, it is true the number of infeasible path conditions remains high, but this is an effect of the large amount of alternative infeasible path conditions generated by the sample program. 

By carrying out a thorough analysis of the alternative path conditions produced by \tardis in the first 30 minutes, on average 15 alternative feasible path conditions are produced.
Nevertheless, despite the feasible path conditions are a small number if compared with the large amount of the infeasible ones, the main result is that \tardis with classifier is able to choose, in the first 30 minutes, two thirds of the feasible alternative path conditions produced ($\sim$10 out of $\sim$15).


\section{Conclusions}
\label{sec:conclusions}
At the state of the art, the problem of generating adequate sets of complete test cases has not been satisfactorily solved yet.
This article proposed an approach that distinctively combines dynamic symbolic execution, search-based testing and machine learning, to efficiently generate thorough class-level test suites.
The approach consists of exploring the path space of the target programs with dynamic symbolic execution, instantiating complete test cases with a genetic search algorithm guided with fitness functions that represent the satisfiability of the symbolic path conditions, and prioritizes the symbolic formulas that more likely correspond to  feasible program paths based on an original classification algorithm trained on the characteristics of the formulas for which it ether succeeds or fails overtime.

From the experiments carried out on the program explained in section \ref{sec:motivating}, we can say that this is a successful approach. In particular, the core technical novelty of this approach, i.e. the path selection strategy based on classification algorithm, is effective as it increases the number of feasible paths of the program analyzed and consequently it increases the number of test cases produced.
In the future we are planning to carry out further experiments to confirm the results in a more generalized way by carrying out an extensive experimentation with real programs.

	\bibliographystyle{IEEEtran}
	\bibliography{main}
	
\end{document}